\documentclass[epsfig,12pt]{article}
\usepackage{amssymb}
\usepackage{amsfonts}
\usepackage{graphicx}
\usepackage{amsmath}

\input epsf.sty
\newtheorem{theorem}{Theorem}
\newtheorem{acknowledgement}[theorem]{Acknowledgement}

\input{tcilatex}
\makeatletter \@addtoreset{equation}{section}

\def\be{\begin{equation}}
\def\ee{\end{equation}}
\def\bea{\begin{eqnarray}}
\def\eea{\end{eqnarray}}

\newcommand{\nc}{\newcommand}
\nc{\al}{\alpha} \nc{\bib}{\bibitem} \nc{\la}{\lambda}
\nc{\C}{\mbox{\hspace{1.24mm}\rule{0.2mm}{2.5mm}\hspace{-2.7mm}
C}} \nc{\R}{\mbox{\hspace{.04mm}\rule{0.2mm}{2.8mm}\hspace{-1.5mm}
R}}

\begin{document}

\title{%
\rightline{\mbox {\normalsize
{Lab/UFR-HEP0602/GNPHE/0602/VACBT/0602}}\bigskip}\textbf{\ On Local Calabi-Yau
Supermanifolds and Their Mirrors}}
\author{ R. Ahl Laamara$^{1,2}$, A. Belhaj$^{3,}$\thanks{\texttt{%
abelhaj@uottawa.ca}}\ ,\ \ L.B. Drissi$^{1,2}$\ ,\ E.H. Saidi$^{1,2,}$%
\thanks{\texttt{esaidi@ictp.it}} \\
\textit{{\small $^{1}$ Lab/ UFR de Physique des Hautes Energies}}\\
\textit{\small Facult\'e des Sciences, Rabat, Maroc }\\
\textit{\small $^2$ Virtual African Center for Basic Sciences and Technology
(VACBT)}\\
\textit{\small Groupement National de Physique des Hautes Energies (GNPHE)}\\
\textit{\small Facult\'e des Sciences, Rabat, Maroc }\\
\textit{\small $^3$ Department of Mathematics and Statistics, University of
Ottawa}\\
\textit{\small 585 King Edward Ave., Ottawa, ON, Canada, K1N 6N5 }\\
}
\maketitle

\begin{abstract}
We use local mirror symmetry to study a class of local Calabi-Yau
super-manifolds with bosonic sub-variety \textrm{V}$_{b}$ having a vanishing
first Chern class. Solving the usual super- CY condition, requiring the
equality of the total $U\left( 1\right) $ gauge charges of bosons $\Phi _{b}$
and the ghost like fields $\Psi _{f}$ one $\sum_{b}q_{b}=\sum_{f}Q_{f}$, as $%
\sum_{b}q_{b}=0$ and $\sum_{f}Q_{f}=0$, several examples are studied and
explicit results are given for local $A_{r}$ super-geometries. A comment on
purely fermionic super-CY manifolds corresponding to the special case where $%
q_{b}=0$, $\forall b$ and $\sum_{f}Q_{f}=0$ is also made.\bigskip

\textbf{Key words}: Mirror Symmetry, Local CY super-manifolds, Topological
string theory.
\end{abstract}
\newpage
\tableofcontents

\newpage

\section{Introduction}

Mirror symmetry has played a crucial role in superstring dualities. It
provides a map between Calabi-Yau (CY) manifolds used in the
compactification of 10D superstring models and topological string theory
\cite{FHSV, AKMV}. In particular, it has been shown that the topological A-
and B-models are connected by mirror symmetry \cite{AKMV}; see also
discussion below. However, it has been realized, though, that rigid CY
manifolds can have mirror manifolds which are not themselves CY geometries.
An intriguing remedy is the introduction of CY \emph{super-manifolds} in
these considerations \cite{Se,Sc}. It has thus been suggested that mirror
symmetry is between super-manifolds and manifolds alike, and not just
between bosonic manifolds.

On the other hand, it has been found that there is a correspondence between
the moduli space of holomorphic Chern-Simons theory on the CY super-manifold
$\mathbf{CP}^{3|4}$ and, self-dual, four-dimensional $N=4$ Yang-Mills theory
\cite{W1,PS}. This may also be related to the B-model of open topological
string theory having $\mathbf{CP}^{3|4}$ as target space. Partly based on
this work, CY super-manifolds and their mirrors have subsequently attracted
a great deal of attention \cite{NV}-\cite{R}. It has been found, for
instance, that an A-model defined on the CY super-manifold $\mathbf{CP}%
^{3|4} $ is a mirror of a B-model on a quadric hypersurface in $\mathbf{CP}%
^{3|3}\times \mathbf{CP}^{3|3}$, provided the K\"{a}hler parameter of $%
\mathbf{CP}^{3|4}$ approaches infinity \cite{NV,AV1}.

Following this observation, an effort has been devoted to go beyond these
particular geometries. A special interest has been given to construct the
mirror of Calabi-Yau super-manifolds whose bosonic parts are compact toric
varieties \cite{BDRSS}. One of the objectives of the present work is to
extend the result of \cite{BDRSS} by considering local Calabi-Yau manifolds
which have been used in type II superstring compactifications in presence of
D-branes. In particular, we discuss the mirror symmetry of the topological
A-model on super-manifolds whose the bosonic part is a local CY variety. The
corresponding theory is a supersymmetric $U(1)^{p}$ linear sigma model with $%
(n+p)$ chiral superfields with charge $q_{i}^{a}$ and $2p$ fermionic
superfields with charge given by $Q_{\alpha }^{a}$ which is a $p\times 2p$
matrix. These charges satisfy the super local CY ( SLCY) condition $%
\sum_{i=1}^{p+n}q_{i}^{a}-\sum_{\alpha =1}^{2p}Q_{\alpha }^{a}=0$ requiring
equality between the total charge of bosons and the ghost like fields.

In this paper we shall focus on the mirror super-geometry obtained by first
choosing a special form of the full spectrum of $U(1)^{p}$ gauge charges and
integrating out some fermionic fields in the topological B-model. In this
way, the mirror B-models will still have some fermionic directions. Our
interest will be on the mirror of $ADE$ super-geometries and mainly on the
fermionic extension of the ordinary $A_{r}$ class. First, we study the case
of $A_{1}$ super-geometry, which is found to be closely related to the
equation of the bosonic case in agreement with analysis using LG models.
Ordinary $A_{1}$ geometry is recovered by canceling the fermionic
directions. Then we work out the mirror of a class of $A_{r}$ local super CY
manifold extending the $A_{1}$ super-geometry. Finally, we discuss the
mirror symmetry of local higher dimensional super-CY geometries. In
particular we specialize on the mirror symmetry of the topological A-model
using a fermionic extension of line bundle over $\mathbf{CP}^{n}$.

The organization of this paper is as follows. In section 2, we review mirror
symmetry of local super CY manifolds. In section 3, we study mirror of $ADE$
super-geometries by exhibiting the method on the ordinary $A_{r}$ series. In
section 4, we consider mirror super-geometries beyond $ADE$ and in section
5, we give a conclusion.

\section{Mirror symmetry of local super CY manifolds}

\qquad In this section, we review mirror symmetry for local (bosonic) CY
manifolds \cite{HV,HIV}; then we give the extension to the super case.

\subsection{Bosonic CY}

\qquad To begin, let us consider a two-dimensional $\mathcal{N}=2$
supersymmetric linear sigma model described in terms of $n+p$ chiral
superfields $\Phi _{i}$ with charge $q_{i}^{a},$ $i=1,\ldots
,n+p;\;a=1,\dots ,p$ under $U(1)^{\otimes p}$ gauge symmetry \cite{W2}. The
geometry of the topological A-model can be analyzed by solving the D-term
potential ($D^{a}=0$) of the $\mathcal{N}=2$ linear sigma model; that is
\begin{equation}
\sum\limits_{i=1}^{n+p}q_{i}^{a}|\phi _{i}|^{2}=r^{a},\qquad a=1,\dots ,p,
\label{BDT}
\end{equation}%
where the $r^{a}$'s are FI coupling parameters and where the $\phi _{i}$'s
are the leading scalar fields of the chiral superfield $\Phi _{i}$. Dividing
by $U(1)^{\otimes p}$ gauge symmetry, one gets a $n$-dimensional toric
variety \footnote{%
Note that this geometry can be represented by a toric diagram $\Delta (%
\mathbf{V}^{n})$ spanned by $k=n+p$ vertices $v_{i}$ in a $\mathbb{Z}^{n}$
lattice satisfying $\sum\limits_{i=1}^{n+p}q_{i}^{a}v_{i}=0,\quad \ \ \
a=1,\ldots ,p.$}
\begin{equation}
\mathbf{V}^{n}=\frac{\mathbb{C}^{n+p}\setminus S}{{\mathbb{C}^{\ast }}^{p}},
\label{Vpgen}
\end{equation}%
where the $p$ copies of $\mathbb{C}^{\ast }$ actions indexed by $\
a=1,\ldots ,p$ are given by
\begin{equation}
{\mathbb{C}^{\ast }}^{p}:\phi _{i}\rightarrow \lambda ^{q_{i}^{a}}\phi
_{i},\qquad i=1,\ldots ,n+p,\
\end{equation}%
with $\lambda $ a non zero complex number. The requirement for $\mathbf{V}%
^{n}$ to be a local CY manifold is to impose the condition
\begin{equation}
\sum\limits_{i=1}^{n+p}q_{i}^{a}=0.  \label{cy}
\end{equation}%
On the supersymmetric field theoretic level, this relation implies that the
underlying linear sigma model flows in infrared to a conformal field theory.

Following \cite{HV}-\cite{B}, the mirror B-model is a Landau-Ginzburg (LG)
model with periodic fields $\{Y_{i}\}$ dual to $\{\Phi _{i}\}$ and connected
as,%
\begin{equation}
\func{Re}(Y_{i})=|\Phi _{i}|^{2},\qquad i=1,\ldots ,n+p,
\end{equation}%
where $\func{Re}(Y_{i})$ denotes the real part of $Y_{i}$. Under mirror
transformation, eq(\ref{BDT}) is mapped to
\begin{equation}
\sum_{i}q_{i}^{a}Y_{i}=t^{a},\qquad a=1,...,p,  \label{bmirror}
\end{equation}%
with \ $r^{a}=\func{Re}\left( t^{a}\right) $. Moreover the LG superpotential
of the topological B-model reads as
\begin{equation}
W\left( Y_{1},...,Y_{n+p}\right) =\sum_{i=1}^{n+p}e^{-Y_{i}}.
\label{WmirrorY}
\end{equation}%
For convenience, it is useful to use the following field redefinitions
\begin{equation}
\hat{y}_{i}=e^{-Y_{i}},\qquad i=1,\ldots ,n+p,  \label{yY}
\end{equation}%
Then the superpotential $W=W\left( \hat{y}_{1},...,\hat{y}_{n+p}\right) $
reads as
\begin{equation}
W=\sum_{i=1}^{n+p}\hat{y}_{i},  \label{Wmirrory}
\end{equation}%
and so eq(\ref{bmirror}) translates into the following projective
hypersurface
\begin{equation}
\prod_{i=1}^{n+p}\hat{y}_{i}^{q_{i}^{a}}=e^{-t_{a}},\qquad a=1,...,p,
\label{yc}
\end{equation}%
with the manifest projective symmetry $\hat{y}_{i}\rightarrow \lambda \hat{y}%
_{i}$ following from the CY condition (\ref{cy}). The solution of the
constraint eqs(\ref{yc}) and projective symmetry defines a $\left(
n+p\right) -p-1-1=n-2$ dimensional toric manifold given by a holomorphic
hypersurface in $\mathbb{C}^{n-1}$
\begin{equation}
F(y_{1},...,y_{n-1})=0.  \label{a1}
\end{equation}%
To recover the right dimension of the original manifold; that is a complex
dimension $n$ local CY manifold, we generally use an ad-hoc trick which
consist to add by hand two extra holomorphic variables $u$ and $v$ combined
in a quadratic form $uv$ and modify previous equation as
\begin{equation}
F(y_{1},...,y_{n-1})=uv.  \label{a2}
\end{equation}%
The main objective in what follows is to extend this analysis to a linear
A-model with fermionic (ghosts) fields and study the resulting mirror
B-model. Besides the generalization of above results to local super-CY
manifolds, one of the results following from this fermionic extension is the
re-derivation of eq(\ref{a2}) without need of adding by hand of the term $uv$
of right hand. As we will show later, the new manifold is given by a
hypersurface of type,
\begin{equation}
G(y_{1},...,y_{n-1})=\chi \eta ,
\end{equation}%
where, instead of $u$ and $v$ variables, we have now the variables $\chi $
and $\eta $ which are ghost like fields. As we will see, this relation
defines an even complex $n$ dimension hypersurface of the complex superspace
$\mathbb{C}^{\left( n-1\right) |2}$. This geometry may be then viewed as
alternative elevation of (\ref{a1}). The standard elevation eq(\ref{a2}) is
given by the purely bosonic hypersurface in $\mathbb{C}^{\left( n+1\right)
|0}$.

\subsection{Mirror of local super-CY}

Here we want to study the mirror of the fermionic extension of the
topological A-model on local toric CY manifolds discussed in the previous
subsection. Actually, this may be viewed as an extension of paper \cite%
{BDRSS} which has dealt with the case of compact bosonic toric manifolds.
Important examples of that work have been projective spaces and products
thereof.

\subsubsection{Extended A-model}

Roughly, the extension corresponds to adding, to the usual bosonic
superfield $\Phi _{j}$, a set of $f$-fermionic chiral superfields $\Psi
_{\alpha }$ with $Q_{f}^{a}$ charge under $U(1)^{\otimes p}$ gauge symmetry.
We then have
\begin{eqnarray}
\Phi _{j}\qquad &\rightarrow &\qquad e^{i\sum_{a}\vartheta
_{a}q_{j}^{a}}\Phi _{j},\qquad j=1,...,n+p,  \notag \\
\Psi _{\alpha }\qquad &\rightarrow &\qquad e^{i\sum_{a}\vartheta
_{a}Q_{\alpha }^{a}}\Psi _{\alpha },\qquad \alpha =1,...,f,
\end{eqnarray}%
with same transformations for the leading component fields $\phi _{j}$ and $%
\psi _{\alpha }$ respectively and where the $\vartheta _{a}$'s are the gauge
group parameters. The full spectrum of $U(1)^{\otimes p}$ charge vectors $%
q^{\prime }{}^{a}=\left( q^{a}|Q^{a}\right) $ thus takes the form
\begin{equation}
\left( q^{a}|Q^{a}\right) =\left( q_{1}^{a},\ldots
,q_{p+n}^{a}|Q_{1}^{a},\ldots ,Q_{f}^{a}\right) ,\qquad a=1,\ldots ,p.
\label{qgen}
\end{equation}%
The extended D$^{a}$-term equations resulting from the above generalized
A-model is given by minimizing the Kahler potential of the $2D$ $\mathcal{N}%
=2$ generalized superfield action
\begin{equation}
\mathcal{S}_{\mathcal{N}=2}=\int d^{2}\sigma d^{4}\theta \mathcal{K}+\left(
\int d^{2}\sigma d^{2}\theta \mathrm{W}+cc\right) ,
\end{equation}%
with respect to the gauge superfields $V_{a}$. In above relation, $\mathcal{K%
}$ is the usual gauge invariant Kahler term and $\mathrm{W}$\ is a chiral
superpotential with superfield dependence as,
\begin{eqnarray}
\mathcal{K} &=&\mathcal{K}\left[ \Phi _{1},...,\Phi _{n+p}^{+};\Psi
_{1},...,\Psi _{f}^{+};V_{1},...,V_{p}\right] ,  \notag \\
\mathrm{W} &=&\mathrm{W}\left[ \Phi _{1},...,\Phi _{n+p};\Psi _{1},...,\Psi
_{f}\right] ,
\end{eqnarray}%
as well as coupling constant moduli which have not been specified. Using the
explicit expression of $\mathcal{K}$ and putting back into
\begin{equation}
D^{a}=\frac{\partial \mathcal{K}}{\partial V_{a}}|_{\theta =0}=0,
\end{equation}%
we get the following D$^{a}$-term equations
\begin{equation}
\sum_{i=1}^{p+n}q_{i}^{a}|\phi _{i}|^{2}+\sum_{\alpha =1}^{f}Q_{\alpha
}^{a}|\psi _{\alpha }|^{2}=\func{Re}(t^{a}),\qquad a=1,\ldots ,p,  \label{ED}
\end{equation}%
where $\func{Re}(t^{a})$ stands for FI coupling constant. Strictly speaking,
this is an even hypersurface embedded in the complex supermanifold $\mathbb{C%
}^{p+n|f}$ with dimension $\left( p+n|f\right) $. Therefore, the space of
vacua of above generalized supersymmetric action is a toric super-manifold $%
\mathcal{V}^{n}$ obtained by dividing $\mathbb{C}^{p+n|f}$ by $U(1)^{\otimes
p}$ gauge symmetry group in same logic as in eq(\ref{Vpgen}). Thus we have
\begin{equation}
\mathcal{V}^{n}=\frac{\mathbb{C}^{p+n|f}\setminus \mathrm{S}}{{\mathbb{C}%
^{\ast }}^{p}}.
\end{equation}%
With this relation at hand, one can go ahead and try to develop the toric
super-geometry of these local super-manifolds by mimicking the standard
toric geometry analysis of toric varieties. We shall not do this here; what
we will do rather is to study some specific examples with direct link to
type II superstring theory compactifications. The first class of these
examples concerns specific fermionic extensions of $ADE$ geometries. The
local super-CY (LSCY) condition reads as follows
\begin{equation}
\sum_{i=1}^{p+n}q_{i}^{a}-\sum_{\alpha =1}^{f}Q_{\alpha }^{a}=0.  \label{cyc}
\end{equation}%
This constraint equation is required by the invariance of holomorphic
measure $\mathrm{\Omega }^{\left( p+n|f\right) }$ of the complex superspace $%
\mathbb{C}^{p+n|f}$,
\begin{equation}
\mathrm{\Omega }^{\left( p+n|f\right) }=\left(
\dprod\limits_{i=1}^{n+p}d\phi _{i}\right) \left( \dprod\limits_{\alpha
=1}^{f}d\psi _{\alpha }\right) ,
\end{equation}%
under ${\mathbb{C}^{\ast }}^{p}$ toric symmetry. In what follows, we shall
fix our attention on those local CY super-manifolds $\mathcal{V}^{n}$
obeying the following special solution,%
\begin{equation}
\sum_{i=1}^{p+n}q_{i}^{a}=0,\qquad \sum_{\alpha =1}^{f}Q_{\alpha
}^{a}=0,\qquad a=1,\ldots ,p.  \label{Dgen}
\end{equation}%
In this particular class of solutions of eq(\ref{cyc}), we have taken the
bosonic subvariety as a local CY manifold. This is the case for bosonic
sub-varieties given by the fermionic extensions of $ADE$ geometries we are
interested in here. It is also remarkable that fermionic directions obey as
well a CY condition for bosonic manifold. In conclusion section, we will
make a comment on this issue.

\subsubsection{Extended B-model}

Under T-duality, the bosonic superfield ${\Phi }_{i}$ of the linear
super-toric sigma model is replaced by a dual superfield $Y_{i}$ as before,
while the fermionic superfield ${\Psi }_{\alpha }$ is dualized by the
triplet $(X_{\alpha },{\Large \eta }_{\alpha },{\Large \chi }_{\alpha })$
\cite{AV1}. The bosonic superfields $X_{\alpha }$ is related to $\Psi
_{\alpha }$ as
\begin{equation}
\func{Re}(X_{\alpha })=-|\Psi _{\alpha }|^{2},\qquad \alpha =1,...,f,
\end{equation}%
and the accompanying pair of chiral superfields $\left\{ {\Large \eta }%
_{\alpha }\right\} $ and $\left\{ {\Large \chi }_{\alpha }\right\} $ are
fermionic superfields required by the preservation of the super-dimension
and hence the total central charge. Under this dualization, the original
complex superspace $\mathbb{C}^{p+n|f}$ gets mapped to
\begin{equation}
\mathbb{C}^{p+n+f|2f}.
\end{equation}%
The extended B-model, mirror to the above fermionic extended A-model with
superfield action $\mathcal{S}_{\mathcal{N}=2}$, is given in terms of the
following path integral, see also \cite{BDRSS},
\begin{equation}
\mathcal{Z}=\int \mathcal{D}\digamma \left[ \dprod\limits_{a=1}^{p}{\small %
\delta }\left( \digamma _{a}-t_{a}\right) \right] \exp \left[ \int \mathcal{W%
}\left( Y,X,{\Large \eta },{\Large \chi }\right) \right] ,  \label{ZLG}
\end{equation}%
where we have set $\mathcal{D}\digamma =\left( {\prod_{i}dY_{i}}\right)
\left( {\prod_{\alpha }dX_{\alpha }d{\Large \eta }_{\alpha }d}{\Large \chi }%
_{\alpha }\right) $. In this relation the $\digamma _{a}$'s are the D-terms
of the extended A-model and $\mathcal{W}=\mathcal{W}\left( Y,X,{\Large \eta ,%
}{\Large \chi }\right) $ is the extended LG superpotential of the
topological B-model. They are as follows,%
\begin{eqnarray}
\digamma _{a} &=&\sum_{i=1}^{n+p}q_{i}^{a}Y_{i}-\sum_{\alpha
=1}^{f}Q_{\alpha }^{a}X_{\alpha },\qquad a=1,...,p,  \notag \\
\mathcal{W} &=&\left( \sum_{i=1}^{n+p}e^{-Y_{i}}+\sum_{\alpha
=1}^{f}e^{-X_{\alpha }}(1+{{\Large \eta }_{\alpha }}{\Large \chi }_{\alpha
})\right) .  \label{lg}
\end{eqnarray}%
To extract informations on the local super-geometry of the B-model, we need
to integrate out the delta functions. Below, we shall focus our attention on
the special case where $f=2p$ and exemplify with models which have been used
in type II superstring theory compactifications.

\section{Mirror of $A_{r}$ super-geometries}

Here we focus on the super-geometry extending the usual ordinary $A_{r}$
geometries. A quite similar analysis is a priori possible for the $DE$,
affine and indefinite extensions.

\subsection{Local super A$_{1}$ geometry}

To illustrate the construction, we initially consider the example of the
model $A_{1}$. This is a supersymmetric gauge theory with a $U(1)$ gauge
symmetry and three chiral superfields $\Phi _{i}$ with charge $(1,-2,1)$
together with a real gauge superfield $V$. The D-term constraint (equation
of motion of $V$) reads as
\begin{equation}
|\Phi _{1}|^{2}-2|\Phi _{2}|^{2}+|\Phi _{3}|^{2}=\func{Re}(t).
\end{equation}%
This geometry describes the Kahler deformation of the $A_{1}$ singularity of
the ALE spaces
\begin{equation}
uv=z^{2},
\end{equation}%
where $u,v$ and $z$ are the generators of gauge invariants. They are
realized in terms of the scalar fields as follows
\begin{equation}
u=\Phi _{1}^{2}\Phi _{2},\qquad v=\Phi _{3}^{2}\Phi _{2},\qquad z=\Phi
_{1}\Phi _{2}\Phi _{3}.
\end{equation}%
For generalizations to rank $r\geq 2$ ordinary $ADE$ geometries as well as
affine extensions and beyond see \cite{adil}.

\subsubsection{Extended model}

Basically, there is an abundance of possible fermionic extensions of above
model. It may be limited by imposing the LSCY condition (\ref{cyc}). Since
we are interested in the case $f=2p=2$, the full spectrum of $U(1)$ charge
that one can have is given by the vector
\begin{equation}
q^{\prime }=\left( q|Q\right) =\left( 1,-2,1|1,-1\right) .  \label{qgena}
\end{equation}%
In this construction, $A_{1}$ model appears as a subsystem while, as noted
before and as far as super-CY condition is concerned, there are several
solution of eq(\ref{cyc}). Using the extension (\ref{qgena}), the D-term for
the $A_{1}$ super-geometry becomes
\begin{equation}
|\Phi _{1}|^{2}-2|\Phi _{2}|^{2}+|\Phi _{3}|^{2}+|\Psi _{1}|^{2}-|\Psi
_{2}|^{2}=\func{Re}(t),
\end{equation}%
where $\func{Re}(t)$ is the unique Kahler parameter of the model.

\subsubsection{Mirror of extended model}

Applying mirror transformation to above extended A-model with $A_{1}$
super-geometry, the associated mirror B-model is obtained in the same way as
presented in previous subsection. The corresponding extended LG path
integral eqs(\ref{ZLG}-\ref{lg}) takes the following form
\begin{eqnarray}
\mathcal{Z} &=&\int \mathcal{D}\digamma \delta \left[
Y_{1}-2Y_{2}+Y_{3}-X_{1}+X_{2}-t\right]   \notag \\
&&\times \exp \left( \sum_{i=1}^{3}e^{-Y_{i}}+\sum_{\alpha
=1}^{2}e^{-X_{\alpha }}(1+\eta _{\alpha }\chi _{\alpha })\right) .
\label{Zgen0}
\end{eqnarray}%
with $\mathcal{D}\digamma =\left( \prod_{i=1}^{3}dY_{i}\right) \left(
\prod_{\alpha =1}^{2}dX_{\alpha }d\eta _{\alpha }d\chi _{\alpha }\right) $.
As usually, to extract information on the mirror super-geometry of the
B-model, we integrate out the fermionic fields $\eta _{1},\chi _{1}$. Then
solving the delta function constraint by integrating out $X_{1}$ yields,
\begin{equation}
\mathcal{Z}=\int \mathcal{D}\widetilde{\digamma }\left(
e^{-Y_{1}+2Y_{2}-Y_{3}}e^{-X_{2}}\right) \exp \left( \sum_{i=1}^{3}{\small e}%
^{-Y_{i}}{\small +e}^{-X_{2}}\left[ {\small 1+\eta }_{2}{\small \chi }_{2}%
{\small +e}^{t}{\small e}^{-Y_{1}+2Y_{2}-Y_{3}}\right] \right) ,
\label{Zgen2}
\end{equation}%
where $\mathcal{D}\widetilde{\digamma }=\left( \prod_{i=1}^{3}dY_{i}\right)
\left( dX_{2}d\eta _{2}d\chi _{2}\right) $. Now, introducing new complex
variables $x_{i}$ and $y_{i}$ such that
\begin{equation}
x=e^{-X_{2}},\qquad y_{i}=e^{-Y_{i}},\qquad i=1,2,3,
\end{equation}%
the above partition function becomes
\begin{equation}
\mathcal{Z}=\int \left( dxd\eta _{2}d\chi _{2}\right) \prod_{i=1}^{3}\left(
\frac{dy_{i}}{y_{2}^{3}}\right) \exp \left( \sum_{i=1}^{3}y_{i}+x\left[
1+\eta _{2}\chi _{2}+e^{t}\frac{y_{1}y_{3}}{y_{2}^{2}}\right] \right) .
\label{Zgen}
\end{equation}%
The rescaling $\tilde{x}=\left( x/y_{2}^{3}\right) $ allows us to rewrite
the above path integral as follows
\begin{equation}
\mathcal{Z}=\int dy_{1}dy_{2}dy_{3}d\tilde{x}d\eta _{2}d\chi _{2}\exp \left(
\sum_{i=1}^{3}y_{i}+\tilde{x}y_{2}^{3}\left[ 1+\eta _{2}\chi _{2}+e^{t}\frac{%
y_{1}y_{3}}{y_{2}^{2}}\right] \right) .  \label{Zgen3}
\end{equation}%
In order to get the mirror of local super-geometry $A_{1}$, we can see $%
\tilde{x}$ as a Lagrange multiplier; integrating it out one gets the
following equation of motion
\begin{equation}
1+\eta _{2}\chi _{2}+\frac{y_{1}y_{3}}{y_{2}^{2}}e^{t}=0.  \label{msuperA1}
\end{equation}%
The objective now is to interpret this equation as the mirror constraint
equation of the topological A-model on $A_{1}$ super-geometry. In fact, we
can solve (\ref{msuperA1}) as
\begin{equation}
\frac{y_{1}y_{3}}{y_{2}^{2}}=-(1+\eta _{2}\chi _{2})e^{-t}.
\end{equation}%
Replacing now $t$ by $t^{\prime }=t+i\pi ,$ one absorbs the minus sign
\begin{equation}
\frac{y_{1}y_{3}}{y_{2}^{2}}=e^{-t^{\prime }}+\eta _{2}\chi
_{2}e^{-t^{\prime }}.  \label{cmsuperA1}
\end{equation}%
Actually, this equation is quite similar to the bosonic one; except now we
have the presence of the additional contribution $\eta _{2}\chi
_{2}e^{-t^{\prime }}$ induced by the fermionic fields. It is easy to see
that in the patch $\eta _{2}=\chi _{2}=0$, we recover the bosonic mirror
constraint equation of ALE space with $A_{1}$ singularity, namely
\begin{equation}
\frac{y_{1}y_{3}}{y_{2}^{2}}=e^{-t^{\prime }}.
\end{equation}%
Return to equation (\ref{cmsuperA1}); a straightforward
computation reveals that this equation can be solved by taking the
following parameterization
\begin{eqnarray}
y_{1} &=&y,\qquad y_{3}=\frac{1}{y},  \notag \\
y_{2} &=&(1+\eta _{2}\chi _{2})^{-\frac{1}{2}}e^{\frac{t}{2}}.
\end{eqnarray}%
where we have set $t^{\prime }=t$. We thus end with the following LG
potential
\begin{equation}
y+\frac{1}{y}+(1+\eta _{2}\chi _{2})^{-\frac{1}{2}}e^{\frac{t}{2}}=0,
\label{su}
\end{equation}%
which is mirror to sigma model on $A_{1}$ super-geometry. This equation has
the three following remarkable features:\newline
\textbf{1}. For $\eta _{2}=\chi _{2}$, we recover the usual bosonic LG
superpotential mirror to the bosonic $A_{1}$ geometry
\begin{equation}
y+\frac{1}{y}+e^{\frac{t}{2}}=0.
\end{equation}%
\textbf{2}. In the case $t\rightarrow 0$, one discovers the rule to define
the super extension of the $A_{1}$ singularity with $U\left( 1\right) $
charges as in eq(\ref{qgena}). The mirror of super $A_{1}$ singularity can
be then defined as follows
\begin{equation}
y_{1}y_{3}=y_{2}^{2}(1+\eta _{2}\chi _{2})
\end{equation}%
in agreement with indication from conformal Landau Ginzburg field models
where adjunction of quadratic terms do not modify the total central charge.
Moreover, by using fermionic statistics which forbids higher powers in $\eta
_{2}$ and $\chi _{2}$, one may define extensions of above $A_{1}$
singularity. \newline
\textbf{3}. In the limit where the condensate modulus $\eta _{2}\chi _{2}$
is small, eq(\ref{su}) reduces to
\begin{equation}
y+\frac{1}{y}+e^{\frac{t}{2}}-\frac{1}{2}\eta _{2}\chi _{2}e^{\frac{t}{2}}=0.
\end{equation}%
By making the identification $\frac{1}{2}\eta _{2}\chi _{2}e^{-\frac{t}{2}}$
with the $uv$ of the relation (\ref{a1}-\ref{a2}), one discovers that the $uv
$ term added by hand in the bosonic case to recover the right dimension of
mirror manifold, is generated in a natural way in super-geometry. In the
limit $t\rightarrow 0$, we have
\begin{equation}
y+\frac{1}{y}+1=\eta _{2}^{\prime }\chi _{2}^{\prime },  \label{superA11}
\end{equation}%
where we have set $\eta _{2}^{\prime }\chi _{2}^{\prime }=\frac{1}{2}\eta
_{2}\chi _{2}$. This is a complex two dimensions even hypersurface of $%
\mathbb{C}^{1|2}$.

\subsection{Super $A_{p}$}

\qquad Now, we would like to push further the above\ results on super $A_{1}$
to the class of $A_{p}$ super-geometry series having usual $A_{p}$ geometry
as local bosonic Calabi-Yau submanifolds. To start, recall that $A_{p}$
geometry has a description in terms of $U(1)^{\otimes p}$ sigma model
involving $\left( p+2\right) $ chiral fields with the bosonic charge $%
p\times \left( p+2\right) $ matrix
\begin{equation}
q_{i}^{a}=\left(
\begin{array}{rrrrrrrrrrr}
1 & -2 & 1 & 0 & 0 & \dots  & 0 & 0 & 0 & 0 & 0 \\
0 & 1 & -2 & 1 & 0 & \dots  & 0 & 0 & 0 & 0 & 0 \\
0 & 0 & 1 & -2 & 1 & \dots  & 0 & 0 & 0 & 0 & 0 \\
\vdots  &  &  &  &  &  &  &  &  &  & \vdots  \\
0 & 0 & 0 & 0 & 0 & \dots  & 0 & 0 & 1 & -2 & 1%
\end{array}%
\right) .
\end{equation}%
Basically, there are several fermionic extensions of the above A-model.
However as we mentioned before, we consider a model with $2p$ fermionic
fields. In this way, the SLCY condition may limit the choice of the charge
matrix. For a reason to be specified later on, we propose the following $%
U(1)^{\otimes p}$ charge spectrum for ghost like superfields
\begin{equation}
Q_{\alpha }^{a}=\left(
\begin{array}{rrrrrrrrr}
1 & -1 & 0 & 0 & \dots  & 0 & 0 & 0 & 0 \\
0 & 0 & 1 & -1 & \dots  & 0 & 0 & 0 & 0 \\
0 & 0 & 0 & 0 & \dots  & 0 & 0 & 0 & 0 \\
\vdots  &  &  &  &  &  &  &  & \vdots  \\
0 & 0 & 0 & 0 & \dots  & 0 & 0 & 1 & -1%
\end{array}%
\right) .
\end{equation}%
This representation constitues a simple and natural extension of eq(\ref%
{qgena}) recovering $A_{1}$ super-geometry as a leading example; other
representations are obviously possible. This choice of $U(1)^{\otimes p}$
charge matrix for ghost like fields allows us to handle each line as an
individual $A_{1}$ super-geometry. In this way, we can easily repeat the
same lines that we have done for the super $A_{1}$ case. Let us give some
details below.

Roughly, LG mirror superpotential is given in terms of the following path
integral
\begin{equation}
\mathcal{Z}=\int \mathcal{D}\digamma \left[ \dprod\limits_{a=1}^{p}\delta
\left( \digamma _{a}-t_{a}\right) \right] \exp \left( \sum_{i}{\small e}%
^{-Y_{i}}{\small +}\sum_{\alpha }{\small e}^{-X_{\alpha }}{\small (1+\eta }%
_{\alpha }{\small \chi }_{\alpha }{\small )}\right) ,
\end{equation}%
where now $\mathcal{D}\digamma =\left( {\prod_{i=1}^{p+2}dY_{i}}\right)
\left( {\prod_{\alpha =1}^{2p}dX_{\alpha }d\eta _{\alpha }d\chi _{\alpha }}%
\right) $ and where we have set
\begin{equation}
\digamma _{a}=Y_{a}-2Y_{a+1}+Y_{a+2}-X_{2a-1}+X_{2a}.
\end{equation}%
This partition function $\mathcal{Z}$ has $p$ delta functions $\delta \left(
\digamma _{a}-t_{a}\right) $. To get the mirror super-geometry, we first
integrate out the fermionic field variables $(\eta _{1a}\chi _{1a})$ leaving
only a dependence on $(\eta _{2a}\chi _{2a}),$ $a=1,2,\dots ,p$ and then we
use delta functions to eliminate the field variables $X_{2a-1}$. In doing so
and following the same way as before, we get $p$ equations of motion,
\begin{equation}
1+\eta _{2a}\chi _{2a}=\prod_{i}{y}_{i}^{q_{i}^{a}},\qquad a=1,\ldots ,p.
\end{equation}%
To see how to obtain these equations, let us consider the case of $A_{2}$
super-geometry. This is a $U(1)^{2}$ gauge theory with four chiral
superfields $(\Phi _{1},\Phi _{2},\Phi _{3},\Phi _{4})$ and four ghost like
ones $(\Psi _{1},\Psi _{2},\Psi _{3},\Psi _{4})$. The full spectrum of $%
U(1)^{2}$ gauge charges is given by
\begin{eqnarray}
q^{\prime }{}^{1} &=&\left( 1,-2,1,0|1,-1,0,0\right) ,  \notag \\
q^{\prime }{}^{2} &=&\left( 0,1,-2,1|0,0,1,-1\right) .
\end{eqnarray}%
The above path integral reduces in present case to
\begin{equation}
\mathcal{Z}=\int \mathcal{D}\digamma \delta \left( \digamma
_{1}-t_{1}\right) \delta \left( \digamma _{2}-t_{2}\right) \exp \left(
\sum_{i=1}^{4}e^{-Y_{i}}+\sum_{\alpha =1}^{4}e^{-X_{\alpha }}(1+\eta
_{\alpha }\chi _{\alpha })\right) ,  \label{ZLG2}
\end{equation}%
with field measure $\mathcal{D}\digamma =\left( {\prod_{i=1}^{4}dY_{i}}%
\right) \left( {\prod_{\alpha =1}^{4}dX_{\alpha }d\eta _{\alpha }d\chi
_{\alpha }}\right) $ and D-terms as
\begin{eqnarray}
\digamma _{1} &=&Y_{1}-2Y_{2}+Y_{3}-X_{1}+X_{2},  \notag \\
\digamma _{2} &=&Y_{2}-2Y_{3}+Y_{4}-X_{3}+X_{4}.
\end{eqnarray}%
Integrating in similar way as we have done for $A_{1}$ super-geometry and
making the same variable changes, we get
\begin{eqnarray}
\mathcal{Z} &=&\int \mathcal{D}\digamma ^{\prime }\exp \left[
\sum_{i=1}^{4}y_{i}+\tilde{x_{1}}y_{2}^{2}\left( 1+\eta _{2}\chi _{2}+\frac{%
e^{t_{1}}y_{1}y_{3}}{y_{2}^{2}}\right) \right]   \notag \\
&&\times \exp \left[ \tilde{x_{2}}y_{3}^{2}\left( 1+\eta _{4}\chi _{4}+\frac{%
e^{t_{2}}y_{2}y_{4}}{y_{3}^{2}}\right) \right] .  \label{Zgen4}
\end{eqnarray}%
with $\mathcal{D}\digamma ^{\prime }=\left( \prod_{i=1}^{4}dy_{i}\right)
\left( d\tilde{x_{1}}d\tilde{x_{2}}d\eta _{2}d\chi _{2}d\eta _{4}d\chi
_{4}\right) $. In this case, we have two equations of motion which are given
by
\begin{eqnarray}
\frac{y_{1}y_{3}}{y_{2}^{2}} &=&\left( 1+\eta _{2}\chi _{2}\right)
e^{-t_{1}^{\prime }},  \notag  \label{cmsuperA2} \\
\frac{y_{2}y_{4}}{y_{3}^{2}} &=&\left( 1+\eta _{4}\chi _{4}\right)
e^{-t_{2}^{\prime }},
\end{eqnarray}%
with $t_{a}^{\prime }=t_{a}+i\pi $. After solving these two equations, we
come up with the following mirror relation%
\begin{equation}
\frac{1}{y}+\left( e^{\frac{t_{1}^{\prime }}{2}}({1+\eta _{2}\chi _{2}}%
)\right) +y+y^{2}\left( e^{-t_{2}^{\prime }}e^{\frac{-t_{1}^{\prime }}{2}}{%
(1+\eta _{4}\chi _{4})(1+\eta _{2}\chi _{2})}^{\frac{1}{2}}\right) =0,
\end{equation}%
which should be compared with the usual mirror relation of ordinary $A_{2}$
geometry $\frac{1}{y}+1+y+y^{2}=0$ associated to the limit $t_{a}^{\prime }=0
$ and ${\eta _{2}=\chi _{2}=0}$.

\section{More on mirror super-geometry}

\qquad The method developed so far can be also used to build other local
super CY manifolds. A simple extension of above $A_{1}$ super-geometry
analysis is given by a sigma model with target space involving a fermionic
extension of line bundle over $\mathbf{CP}^{p}$ with $p\geq 2$. The case $p=1
$ corresponds exactly to the $A_{1}$ super-geometry studied previously. Let
us analyze the case $p=2$; that is the line bundle $\mathcal{O}(-3)$ over $%
\mathbf{CP}^{2}$. It admits a $U(1)$ sigma model description in terms of
four bosonic chiral fields with charge vector $(1,1,1,-3)$ and the
corresponding D-term equation is given by
\begin{equation}
|\Phi _{1}|^{2}+|\Phi _{2}|^{2}+|\Phi _{3}|^{2}-3|\Phi _{4}|^{2}=\func{Re}(t)
\end{equation}%
Adding now $2$ ghost like field variables $\Psi _{1}$ and $\Psi _{2}$ with
vector charge $(1,-1),$ as required by SLCY condition, the D-term constraint
equation of the extended A-model is given by
\begin{equation}
|\Phi _{1}|^{2}+|\Phi _{2}|^{2}+|\Phi _{3}|^{2}-3|\Phi _{4}|^{2}+|\Psi
_{1}|^{2}-|\Psi _{2}|^{2}=\func{Re}(t).
\end{equation}%
The corresponding mirror super-geometry is given in terms of the following
path integral
\begin{equation}
\mathcal{Z}=\int \mathcal{D}\digamma \delta \left( \digamma -t\right) \exp
\left( \sum_{i=1}^{4}e^{-Y_{i}}+\sum_{\alpha =1}^{2}e^{-X_{\alpha }}(1+\eta
_{\alpha }\chi _{\alpha })\right) .  \label{Zgen5}
\end{equation}%
with $\mathcal{D}\digamma =\left( \prod_{i=1}^{4}dY_{i}\right) \left(
\prod_{\alpha =1}^{2}dX_{\alpha }d\eta _{\alpha }d\chi _{\alpha }\right) $
and
\begin{equation}
\digamma =Y_{1}+Y_{2}+Y_{3}-3Y_{4}-X_{1}+X_{2}.
\end{equation}
Now integrating out the fermionic fields $\eta _{1},\chi _{1}$ and solving
the delta function constraint by eliminating $X_{1}$, we get
\begin{eqnarray}
\mathcal{Z} &=&\int \mathcal{D}\digamma \left(
e^{-Y_{1}-Y_{2}-Y_{3}+3Y_{4}}e^{-X_{2}}\right)   \notag \\
&&\times \exp \left( \sum_{i=1}^{4}e^{-Y_{i}}+e^{-X_{2}}(1+\eta _{2}\chi
_{2}+e^{t}e^{-Y_{1}-Y_{2}-Y_{3}+3Y_{4}})\right) ,  \label{Zgen7}
\end{eqnarray}%
with $\mathcal{D}\digamma =\left( \prod_{i=1}^{4}dY_{i}\right) \left(
dX_{2}d\eta _{2}d\chi _{2}\right) $. Using the following field re-definition
\begin{equation}
y_{i}=e^{-Y_{i}},\qquad x=e^{-X_{2}}
\end{equation}%
the above equation becomes
\begin{equation}
\mathcal{Z}=\int \left( \prod_{i=1}^{4}\frac{dy_{i}}{y_{4}^{4}}\right)
\left( dxd\eta _{2}d\chi _{2}\right) \exp \left[ \sum_{i=1}^{4}y_{i}+x\left(
1+\eta _{2}\chi _{2}+\frac{e^{t}y_{1}y_{2}y_{3}}{y_{4}^{3}}\right) \right] .
\label{Zgen8}
\end{equation}%
By help of the following rescaling $\tilde{x}=\frac{x}{y_{4}^{4}}$, the
mirror geometry becomes
\begin{equation}
\mathcal{Z}=\int d\tilde{x}d\eta _{2}d\chi _{2}\prod_{i=1}^{4}dy_{i}\exp %
\left[ \sum_{i=1}^{4}y_{i}+\tilde{x}y_{4}^{4}\left( 1+\eta _{2}\chi _{2}+%
\frac{e^{t}y_{1}y_{2}y_{3}}{y_{4}^{3}}\right) \right] .  \label{Zgen9}
\end{equation}%
In this case, the equation of motion reads as
\begin{equation}
\frac{y_{1}y_{2}y_{3}}{y_{4}^{3}}=-(1+\eta _{2}\chi _{2})e^{-t}.
\end{equation}%
Absorbing the minus sign by replacing $t$ by $t+i\pi $, the above equation
becomes
\begin{equation}
\frac{y_{1}y_{2}y_{3}}{y_{4}^{3}}=e^{-t^{\prime }}+\eta _{2}\chi
_{2}e^{-t^{\prime }}.
\end{equation}%
This can be easily solved by the following parameterization
\begin{eqnarray}
y_{1} &=&x,\qquad y_{2}=y,\qquad y_{3}=\frac{1}{xy}  \notag \\
y_{4} &=&(1+\eta _{2}\chi _{2})^{\frac{-1}{3}}e^{\frac{t^{\prime }}{3}}.
\end{eqnarray}%
The superpotential describing the mirror of the super-geometry reads as
\begin{equation}
x+y+\frac{1}{xy}+(1+\eta _{2}\chi _{2})^{\frac{-1}{3}}e^{\frac{t^{\prime }}{3%
}}=0.
\end{equation}%
For $\eta _{2}\chi _{2}=0$, we rediscover the usual bosonic relation.

\section{Conclusion}

\qquad In this paper, we have studied mirror symmetry of A-model on
Calabi-Yau super-manifolds constructed as fermionic extensions of local
toric CY satisfying the SLCY condition $\sum_{i=1}^{p+n}q_{i}^{a}=\sum_{%
\alpha =1}^{2p}Q_{\alpha }^{a}$. By solving this condition as $%
\sum_{i=1}^{p+n}q_{i}^{a}=0$ and $\sum_{\alpha =1}^{2p}Q_{\alpha }^{a}=0$
separately; we have considered two classes of mirror super-geometries. The
first class deals with a special fermionic extension of ordinary geometries
and the second class concerns a set of sigma models with target space
involving a fermionic extension of line bundle over $\mathbf{CP}^{n}$ with $%
n\geq 2$. The representations studied here are not the general
ones since the bosonic subvariety of super-manifold considered
here is taken as a Calabi-Yau manifold. This condition is
obviously not a necessary condition for building Calabi-Yau
super-manifolds. This work may be viewed as a extension of
\cite{BDRSS} which has dealt with bosonic compact toric varieties.
The mirror geometries studied in that paper have dealt only with
bosonic variables. However, here the mirror B-models involve
fermionic
directions captured by the ghost like fields. In dealing with the mirror of $%
A_{r}$ super-geometries, we have shown that these local CY super-manifolds
are described by algebraic geometry equations quite similar to the bosonic
case. The later can be obtained by canceling fermionic directions. Moreover,
we have found that in super-geometry, the right dimension of the bosonic CY
subvariety is recovered in natural as shown on (\ref{superA11}). Finally we
have shown that this approach applies as well to higher dimensional mirror
super-geometries; mirror of A-model on super line bundle over $\mathbf{CP}%
^{n}$ studied in section 4 is an example amongst others.

\qquad In the end of this study, we would like to add that along with
ordinary CY manifolds embedded in $\mathbb{C}^{n|0}$ and super-CY manifolds
embedded in complex space $\mathbb{C}^{n|m}$, we may also have super CY
manifolds embedded in the purely fermionic space $\mathbb{C}^{0|m}$ without
basic bosonic coordinates. These special super CY varieties are then
hypersurface in $\mathbb{C}^{0|m}$ involving ghost fields like only.

\begin{acknowledgement}
:\qquad\ \newline
AB would like to thank C. Vafa for kind hospitality at Harvard University,
where a part of this work is done. We thank J. Rasmussen and A. Sebbar for
collaborations related to this work. This research work is supported by the
program Protars III D12/25, CNRST.
\end{acknowledgement}

\end{document}